\documentclass[fleqn,usenatbib]{mnras}

\usepackage{newtxtext,newtxmath}
\usepackage[T1]{fontenc}

\DeclareRobustCommand{\VAN}[3]{#2}
\let\VANthebibliography\thebibliography
\def\thebibliography{\DeclareRobustCommand{\VAN}[3]{##3}\VANthebibliography}

\usepackage{graphicx}	% Including figure files
\usepackage{amsmath}	% Advanced maths commands

\title[Constraining CE evolution in BNS formation]{Constraining Common Envelope Evolution in Binary Neutron Star Formation with Combined Galactic and Gravitational-Wave Observations}

\author[Chen et al.]{
Zhiwei Chen,$^{1,2}$
Jihui Zhang,$^{1,2}$
Youjun Lu,$^{1,2}$\thanks{E-mail: luyj@nao.cas.cn}
Jifeng Liu, $^{1,2}$ \thanks{E-mail:  jfliu@nao.cas.cn}
Changwen Zeng, $^{1,2}$
\\
$^{1}$National Astronomical Observatories, Chinese Academy of Sciences, 20A Datun Road, Beijing 100101, China\\
$^{2}$School of Astronomy and Space Sciences, University of Chinese Academy of Sciences, 19A Yuquan Road, Beijing 100049, China\\
}

\date{Accepted XXX. Received YYY; in original form ZZZ}

\pubyear{\the\year{}}

\begin{document}
\label{firstpage}
\pagerange{\pageref{firstpage}--\pageref{lastpage}}
\maketitle

\begin{abstract}
Binary neutron stars (BNSs) are among the most interesting sources for multimessenger studies. A number of recently discovered BNSs in the Milky Way by radio telescopes have added new information to the parameter distribution of the Galactic BNSs. The {scarcity} of BNS mergers during the O4 run of the LIGO-Virgo-Kagra (LVK) suggests a BNS local merger rate six times lower than the previous constraint obtained by O1-O3 runs. With these new multimessenger observations, in this letter, we adopt the compact binary population synthesis model and Bayesian analysis to constrain the formation and evolution of BNSs, especially the common envelope (CE) evolution. We find that it is required: (1) a fraction ($f_{\rm HG}\sim0.8$) but not all of the Hertzsprung gap donors merged with their companions in the CE stage, in order to simultaneously explain the low BNS merger rate density and the existence of the short-orbital-period ($\lesssim 1$\,day) Galactic BNSs, different from either all ($f_{\rm HG}=1$) or none ($f_{\rm HG}=0$) adopted in previous studies; (2) a large CE ejection efficiency $\alpha$ ($\sim 5$), in order to explain the existence of the long-orbital-period ($\gtrsim 10$\,day) Galactic BNSs.  
\end{abstract}


%%%%%%%%%%%%%%%%%%%%%%%%%%%%%%%%%%%%%%%%%%%%%%%%%%

%%%%%%%%%%%%%%%%% BODY OF PAPER %%%%%%%%%%%%%%%%%%
\section{Introduction}

The discovery of the first
gravitational wave (GW) emitted by the binary neutron star (BNS) merger, i.e.,  GW170817 \citep{2017PhRvL.119p1101A} together with its various electromagnetic (EM) counterparts \citep[e.g., ][]{2017ApJ...848L..12A, 2017ApJ...848L..13A,2017Sci...358.1556C, 2019MNRAS.489L..91C, 2019PhRvX...9a1001A}  mark the beginning of the new era of multi-messenger astrophysics, leading to a more extensive investigation on the formation and evolution of BNS systems since the discovery of the Hulse-Taylor pulsar \citep{1975ApJ...195L..51H}. 
There are mainly two formation channels for BNSs, i.e., isolated evolution of massive binary stars \citep[hereafter BSE channel, e.g.,  ][]{2002MNRAS.329..897H,2016ApJ...819..108B,2019ApJ...880L...8A, 2022LRR....25....1M} and dynamical interaction in dense stellar systems \citep[hereafter dynamical channel, e.g., ][]{2019MNRAS.488...47F, 2020ApJ...888L..10Y,2020ApJ...898..152S}. Notably, NSs are not heavy enough to sink into the center of the globular cluster and the formation via the dynamical channel may be quite inefficient \citep{2016ApJ...819..108B}. Thus, the BSE channel are believed to be the dominant channel for BNS formation.

The compact binary population synthesis (hereafter CBPS) method has been widely adopted \citep[e.g.,][]{2020MNRAS.494.1587C,2022MNRAS.509.1557C,2023ApJS..264...45F,2024ApJ...963...80D,2025arXiv250602316M} to help {understand} the the complex astrophysical processes involved in BSE channel, such as the most uncertain common envelope (CE) evolution \citep[e.g.,][]{2018MNRAS.481.4009V,2018MNRAS.480.2011G,2019MNRAS.486.3213A}. For example, in our previous work \citep[][hereafter CYL22]{2022MNRAS.509.1557C}, we found that models with efficient CE ejection are more supported by the local merger rate density $R_0\sim 320_{-240}^{+390}\rm ~Gpc^{-3}~yr^{-1}$ estimated by O1-O3 runs of LIGO-Virgo-KAGRA \citep{2021arXiv211103606T} and the Galactic BNSs. 

Recent discoveries of new Galactic BNSs by advanced radio telescopes (e.g., \citealt{2024MNRAS.530.1506S,2025RAA....25a4003W}; see also Table~\ref{tab:DNS_params_ordered}) have enriched our knowledge of their population properties. Meanwhile, the O4 run of the LVK collaboration is nearing completion, yielding decisively tighter constraints on the BNS local merger rate density. Recently, a preliminary catalog from \citet{2025arXiv250708778A} reports
$R_0\sim 56^{+99}_{-40}\rm ~Gpc^{-3}~yr^{-1}$ based on O4 public alerts, significantly lower than O1-O3 estimates, indicating that the constraints on the formation and evolution of BNSs should be revisited. In this letter, we revisit such constraints with CBPS model by introducing a new parameter $f_{\rm HG}$ beyond the ejection efficiency $\alpha$, representing the fraction that the Hertzsprung gap donor stars cannot survive from CE and find out the most compatible model with current multimessenger observations. This letter is organized as follows. In section~\ref{sec:bse}, we introduce our methodology on simulations.  In section~\ref{sec:results}, we present our main results on the constraints of the CE evolution. The conclusions and discussions are provided in section~
\ref{sec:con}.

\section{Methodology}

\subsection{BSE Simulation Settings}
\label{sec:bse}

In this work, we simulate the formation and evolution of BNS systems via an updated version of CBPS code \textbf{BSE}, which is originally developed by \citet{2000MNRAS.315..543H,2002MNRAS.329..897H}, and later modified by \citet{2010A&A...521A..85Y} and CYL22. Many important recipes in the binary evolution are take into account,  
here we only give a brief introduction on those updated settings compared with CYL22. 

\textbf{Initial Condition}
The initial mass spectrum for main-sequence (MS) binary stars are generated by assuming that, the primary mass $M_1$ follows $p(M_1)\propto M_1^{-2.5}$ for $M_1>1M_{\odot}$ \citep{2001MNRAS.322..231K}, and the mass ratio $q=M_2/M_1$ follows a uniform distribution within $(0.01,1)$ \citep{1989ApJ...347..998E}. The initial semimajor axes $a$ of binaries are assumed to follow the distribution given by \citet{1998MNRAS.296.1019H}, and the initial eccentricities of the orbits are assumed to be zero.  
As for the newborn NS, we assume the equation of state to be SLy with the maximum mass for non-rotating NSs $M_{\rm TOV}\sim 2.06M_{\odot}$ \citep{2001A&A...380..151D}. In addition, we set the initial spin period $P_0=0.01\rm s$.
The initial magnetic field in unit of Gauss follows a log-normal distribution with mean of $13$ and scatter of $0.55$ for NS with progenitor MS mass $\lesssim 20 M_{\odot}$ \citep[e.g.,][]{2006ApJ...643..332F}, and a uniform distribution within the range of $[10^{14},10^{15}]$ for NS with progenitor MS mass $\gtrsim 20 M_{\odot}$ \citep[e.g.,][]{2014ApJS..212....6O}.

\textbf{Common Envelope}: One of the most uncertain process in binary evolution is the common envelope phase. In this work,  we adopt the simplified but widely-used $\alpha$-formalism of binding energy \citep[e.g.,][]{1984ApJ...277..355W}, i.e., 
\begin{equation}
\label{eq:CE}
    |E_{\rm bind}|=\frac{G(M_1-m_{1c})M_1}{\lambda r_{\rm L}}=\alpha\bigg(\frac{Gm_{1c}M_2}{2a_{\rm f}}-\frac{GM_1M_2}{2a_{\rm i}}\bigg),
\end{equation}
where $\alpha$ is the CE ejection efficiency and $\lambda$ is the structure parameters, which is drawn from \citet{2014A&A...563A..83C} \citep[see also,][]{2021A&A...650A.107M}. The symbol $r_{\rm L}$ represents the Roche lobe radius, $a_{\rm i}$ and $a_{\rm f}$ are the semi-major axis of the binary before and after the CE phase, $M_1$ and $m_{1c}$ are the masses of the star and its core, $M_2$ is the companion mass. The original definition of such $\alpha-$description requires that $\alpha\lesssim 1$, representing that the unbind energy of CE only comes from the orbital energy \citep[e.g.,][]{2023LRCA....9....2R}, but later also extended to the cases of $\alpha\gtrsim 1$ \citep[e.g.,][]{2022MNRAS.509.1557C,2023MNRAS.526.2210S}, supported by that there may be additional energy contribute to the unbind of CE. \citet{2025ApJ...980..181C} demonstrated that models with $\alpha<1$ cannot reproduce the observed Galactic BNSs with long orbital period, because the orbits of their progenitors shrink significantly after the CE stage in these models. We also verify such statement for the case of $\alpha=1$, which can be seen in Table~\ref{table:para}.

More importantly,  the Hertzsprung-gap (HG) donors in CE are worth special attention \citep[e.g.,][]{2007ApJ...662..504B,2012ApJ...759...52D,2013ApJ...779...72D}, since HG stars may have not yet developed a steep density (or entropy) gradient between core and envelope \citep{2004ApJ...601.1058I}, i.e., there is no clear boundary between the envelope and core structure. Therefore, \citet{2007ApJ...662..504B} first proposed that the CE initiated by HG stars may lead to premature merger and hence result no compact binary merger in the end. By such a treatment, \citet{2017MNRAS.472.2422M} found that the predicted merger rate of stellar binary black holes can match with the LVK observations. Otherwise, the rate will be too high to be consistent with GW observations. Therefore, a large number of works utilizing CBPS method have adopted this treatment \citep[e.g.,][also known as ``pessimistic CE"]{2018MNRAS.480.2011G,2020ApJ...898..152S, 2024ApJ...976...24B}. Facing the same situation indicated by the latest constraint on the local BNS merger rate density, i.e., a smaller value $R_0\sim 56^{+99}_{-40}\rm ~Gpc^{-3}~yr^{-1}$ \citep{2025arXiv250708778A}, such treatment may be also needed for BNSs. However, later we will see that if all the HG donors are not allowed to survive from the CE stage, the formation of Galactic BNSs possessing short orbital period are significantly suppressed. Therefore, in this work, we introduce a new parameter $f_{\rm HG}$
in the BSE code and assume that a fraction of $f_{\rm HG}$ HG donors are merged if they enter a CE, while the rest $1-f_{\rm HG}$ can survive from CE stage and evolve continuously.

\textbf{Supernova Explosion}: 
In this work, we identify three different Supernova explosion (SNe) mechanisms, including electron-capture SNe \citep[ECSNe,][]{1980PASJ...32..303M}, ultra-stripped SNe \citep[USSNe,][]{2015MNRAS.451.2123T} and core-collapse SNe \citep[CCSNe,][]{2002MNRAS.329..897H}, by the Helium core masses of progenitor stars and adopt the remnant mass function given by the rapid model from \citet{2012ApJ...749...91F}. As for the natal kick, we assume its orientation to be randomly distributed and its magnitude $v_{\rm k}$ follows a Maxwellian distribution \citep{2005MNRAS.360..974H}, with dispersion $\sigma_{\rm k}$ of $30\rm km/s$ and $265\rm km/s$ for ECSNe/USSNe and CCSNe respectively \citep[e.g.,][]{2005MNRAS.360..974H,2018MNRAS.481.4009V}.

\textbf{Spin evolution}: 
In this work, the spin evolution of both isolated NS and NS in binary systems are taken into account. As for NS in binaries, we treat the first NS with a massive donor as an OB or Be-type high-mass X-ray binary (HMXB) system \citep{1997ApJS..113..367B}.  Then we consider
several important processes affecting the spin evolution of the NS, such as wind-fed accretion and Roche-lobe overflow in different phases. As for isolated NS, we assume it spins down via the dipole radiation \citep{2010MNRAS.404.1081R}. Here we refer the readers to Figure 1 in \citet{2025ApJ...980..181C} for more detailed discussions.

{By the above recipes, we simulate mock BNS templates with BSE models possessing various sets of $\alpha$ and $f_{\rm HG}$ (see Tab.~\ref{table:para}, denoted as models M1, M2, to M22), at the metallicty grid of $\log(Z/Z_{\odot}) =-2, -1.5, -1, -0.5, 0$, where the solar metallicity is $Z_{\odot}=0.02$.}

\subsection{{Bayes factor} of Mock Galactic BNS systems }
\label{sec:MW}
We first simulate survived Galactic BNSs and obtain their orbital period and eccentricity , i.e., $(\log{P_{\rm orb}},e)$ by Monte Carlo simulations, with the star formation history (SFH) of Milky way (MW) bulge and disk drawn from \citet{2003MNRAS.345.1381F} and \citet{2014ApJ...781L..31S}, respectively. We {require} that at least one NS in the BNS system should be radio loud for pulsar detection, i.e., above the ``death-line” \citep[e.g.,][]{1991PhR...203....1B}. 
Then we discretize the $(\log{P_{\rm orb}},e)$ distribution map into small pixels and  calculate the likelihood $L_{i}$ for model $M_{i}$ by, 
\begin{equation}
    \log L_i=\sum_{j,l,k=1} N(\log P_{{\rm orb},j},e_{j}\mid \log{P_{l}},e_{k},\sigma^2_{\log P_{l}},\sigma^2_{e_{k}})\mid_{M_i},
\end{equation}
where the summation is over all $(\log{P_{l}},e_{k})$ pixels  and $26$ observed Galactic BNSs 
in the MW field shown in Table~\ref{tab:BNS}. For each pixel, we adopt a 2D {Gaussian} distribution normalized by the number of mock BNS within it, and the standard deviation are assumed to be $\sigma_{\log P_{l}}=0.3$ and $\sigma_{e_{k}}=0.03$, respectively \citep{2018MNRAS.481.4009V}. Finally, given the same prior information, we  estimate the {Bayes factor} $\log{K_i}$  to measure the relative goodness of each model that matches with the Galactic BNS observation, i.e., 
\begin{equation}
    \log K_i=\log L_i-\log L_{\rm m,6},
\end{equation}
where we have chosen model 6 with $\alpha=3$ and $f_{\rm HG}=0$ as our reference models (the same with the model chosen in \citet{2025ApJ...980..181C}).

\begin{figure*}
\centering
\includegraphics[width=2.0\columnwidth]{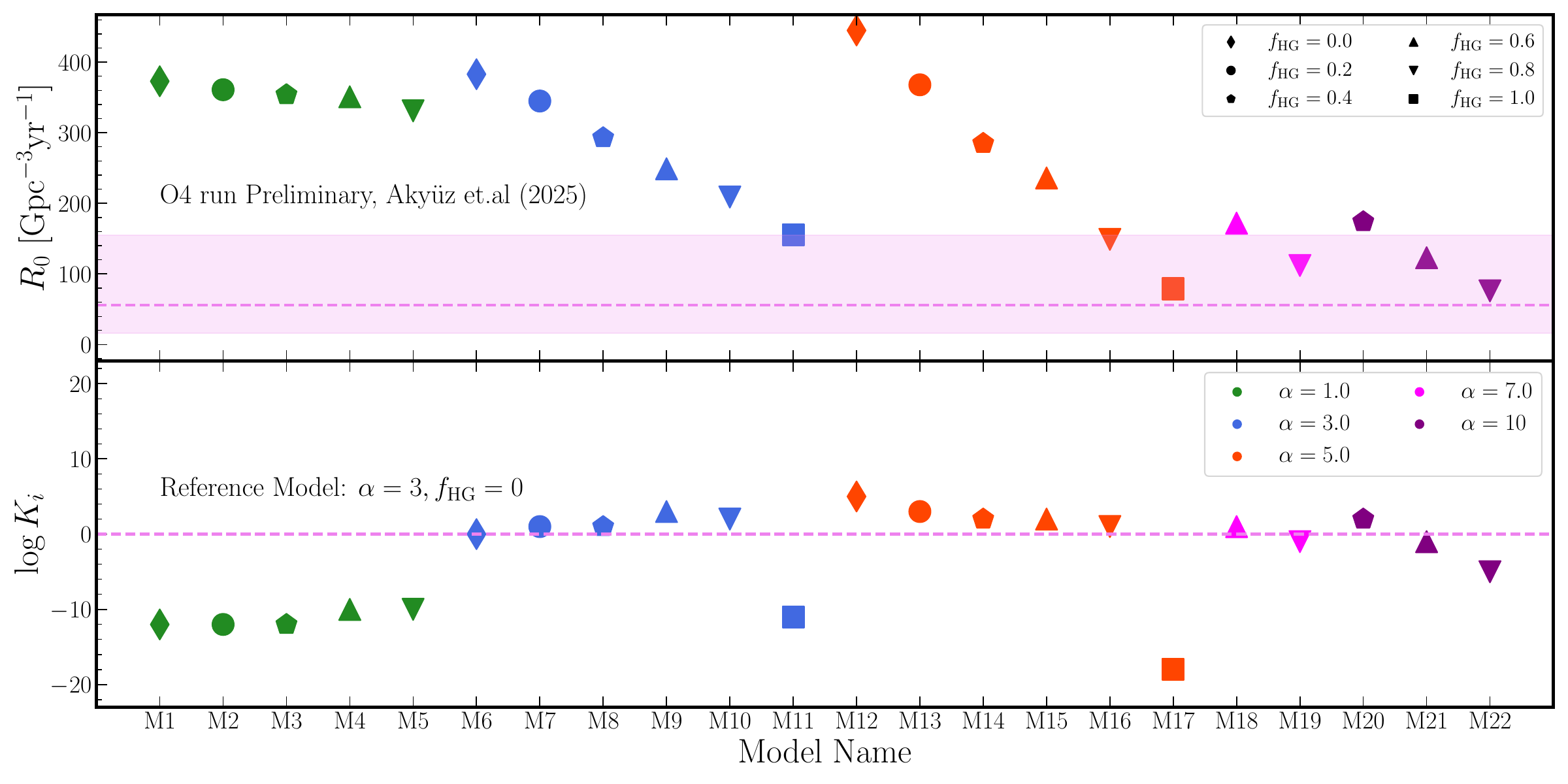}
\caption{The estimated BNS local merger rate density $R_0$ (top panel) and {Bayes factor} of Galactic BNS systems referenced by model 6 with $\alpha=3, f_{\rm HG}=0$ (bottom panel) across models shown in Table~\ref{table:para}. In both panels, the diamond, circle, pentagon, triangle, inverted triangle, and square markers represent models with $f_{\rm HG}=0.0, 0.2, 0.4, 0.6, 0.8, 1.0$ respectively. Green, blue, orange, magenta, and purple color symbols represent models with $\alpha=1.0, 3.0, 5.0, 7.0,$ and $10$, respectively. In the upper panel, the shaded region and pink dashed line show the constraint on the local merger rate density of BNSs from the O4 public alerts by \citet{2025arXiv250708778A}, i.e., $R_0\sim 56^{+99}_{-40}\rm ~Gpc^{-3}~yr^{-1}$. In the lower panel, the pink dashed line show the referenced $\log{K_{i}}=0$ line. 
}
\label{fig:1}
\end{figure*}

\subsection{Local merger rate density}

Based on the templates generated by the BSE simulation, we calculate the BNS local merger rate density by \citep[e.g.,][]{2024ApJ...973..159C}
\begin{equation}
R_0=\int dt_{\rm d}P_t(t_{\rm d}) f_{\rm b}{\rm SFR}(z_{\rm b})\times N_{\rm cor}(Z\left(z_{\rm b})\right),
\label{eq:merger_rate}
\end{equation}
where $f_{\rm b}$ denotes the binary formation efficiency (normally $f_{\rm b}=0.5$), and $t_{\rm d}(z_{\rm b})=\int_{0}^{{z_{\rm b}}}\mid\frac{dt}{dz}\mid dz$ denotes the time delay of a BNS merger from its progenitor binary star formation time and $P_t(t_{\rm d})$ is its probability distribution, ranging from $10\rm Myr$ to a Hubble time $t_{\rm H}$, which can be directly extracted by our simulation. The term ${\rm SFR}(z_{\rm b})$ is the star-formation rate density at the binary formation redshift $z_{\rm b}$ (or time $t_{\rm b}$), and $N_{\rm cor}$ is the number of mergering BNSs per unit mass given a metallicity $Z(z_{\rm b})$ at $z_{\rm b}$. Here in this work, we draw the evolution of ${\rm SFR}(z_{\rm b})$ and $Z(z_{\rm b})$ from \citet{2014ARA&A..52..415M} and \citet{2016Natur.534..512B} respectively.

Below we summarize the key updates introduced in this work relative to CYL22.   
\begin{itemize}
    \item A fraction $f_{\rm HG}$ of HG donors are forced to {merge} with their companion if they enter a CE phase, rather than all systems can survive from CE.
    
    \item The structure parameter $\lambda$ depends on the stage of binary evolution, rather than a constant value. 
    
    \item  Different remnant mass function and kick velocity distribution are considered for different supernova explosion mechanisms, rather than a uniform extrapolating formula.
    
    \item The spin evolution of BNS systems are taking into account by detailed modeling, rather than assuming a constant pulsar lifetime of $10^8\rm yr$. 
    
    \item Different star formation history of local BNS mergers and Galactic BNS systems are taken into account to address their diversity.
    
    \item Nine newly discovered Galactic BNSs and merger rate density estimation from the preliminary O4 catalog are taken into account for constraining BSE parameters.
    
\end{itemize}

\section{Results}
\label{sec:results}

\begin{figure}
\centering
\includegraphics[width=1.0\columnwidth]{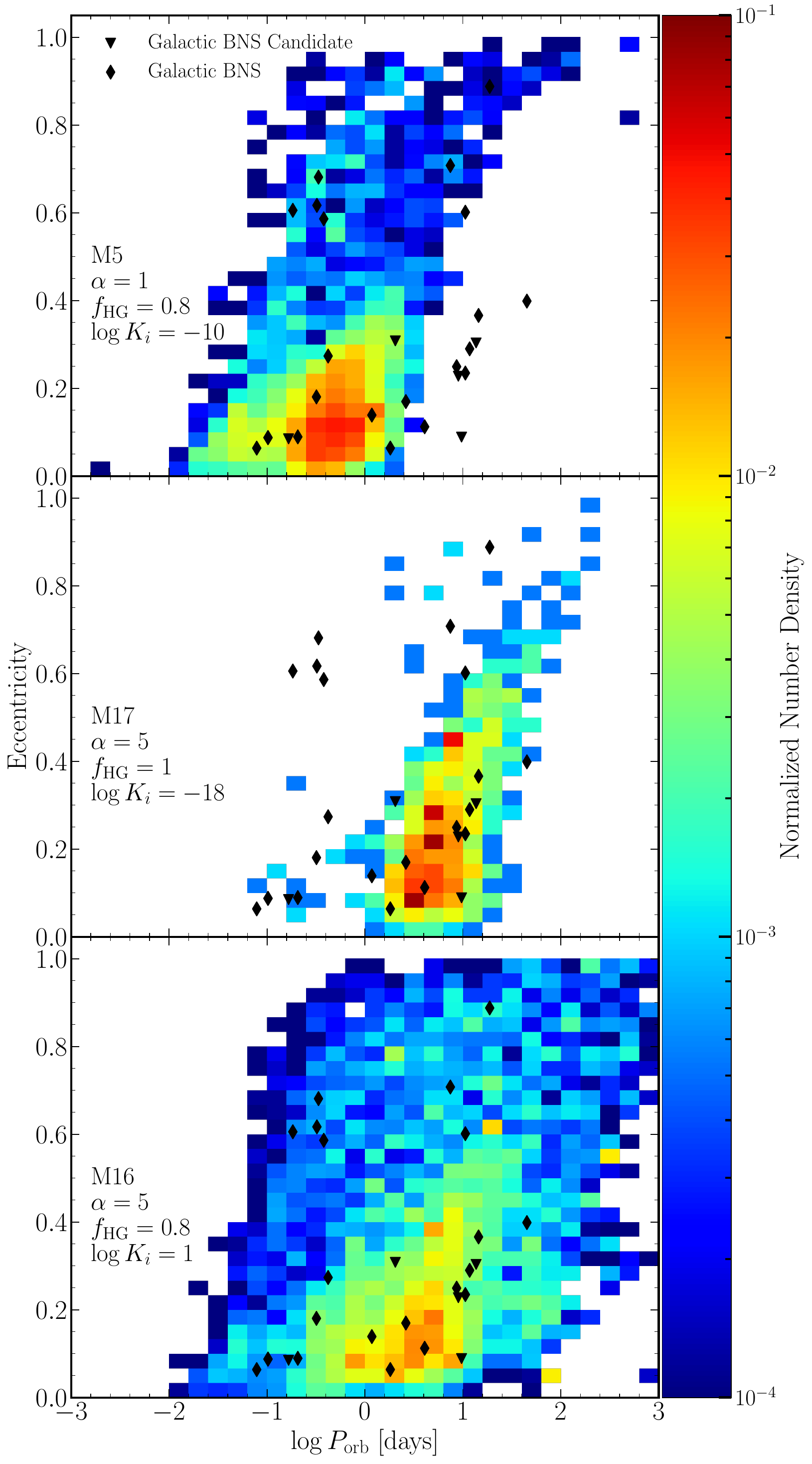}
\caption{The orbital distribution of simulated Galactic BNS systems on the eccentricity-orbital period ($e-\log{P_{\rm orb}}$) plane. The top, middle and bottom panels show the results of model M5 with $\alpha=1, f_{\rm HG}=0.8$, M17 with $\alpha=5, f_{\rm HG}=1$ and M16 with $\alpha=5, f_{\rm HG}=0.8$  respectively. In both panels, the black diamond and triangle show the parameters of  galactic BNS systems and candidates in Table~\ref{tab:DNS_params_ordered}, measured by pulsar radio observation respectively. The colorbar shows the normalized number density of mock BNS systems.
}
\label{fig:2}
\end{figure}

According to the above settings, we obtain mock samples of Galactic BNSs that may be detected by radio telescopes and the local merger rate density for each BSE model listed in Table~\ref{table:para}.
Figure~\ref{fig:1} shows the predicted local BNS merger rate density $R_0$ (top panel) and {Bayes factor} $\log{K_{\rm i}}$ of mock Galactic BNSs (bottom panel) for each BSE model listed in Table~\ref{table:para} with a given set of $\alpha$ and $f_{\rm HG}$. For demonstration, Figure~\ref{fig:2} shows the distributions of mock Galactic BNSs on the $\log{P_{\rm orb}-e}$ plane from models M5, M17, M16, respectively. By careful inspection, 
%we can reach the following results. First, 
first, we find that in the models with $\alpha\lesssim 1$ the formation of Galactic BNSs with long orbital periods, i.e.,  $\log{P_{\rm orb}}\gtrsim 10$\,days is suppressed (see the top panel of Fig.~\ref{fig:2} for model 5). Therefore, these models are disfavored according to their lower {Bayes factor} $\log{K_i}\lesssim -10$, though may produce a BNS merger rate density more consistent with the latest constraint by GW observations. This result is consistent with the findings in  \citet{2025ApJ...980..181C} and can be naturally explained by more severe orbit shrinkage following CE stage at lower $\alpha$ values.

Second, adopting a larger value of $f_{\rm HG}$ leads to significant reduction of $R_0$, as systems with HG donors merging during the CE stage are removed from the potential BNS progenitor population.
For example, for the reference model M6 with $(\alpha,f_{\rm HG})=(3,0)$, the predicted $R_0$ is $\sim 383~\rm Gpc^{-3}yr^{-1}$, while $ R_0\sim 155~\rm Gpc^{-3}yr^{-1}$ for the model M11 with $(\alpha,f_{\rm HG})=(3,1)$. In addition, we observe that if $f_{\rm HG} \geq 0.2$, $R_0$ decreases with larger value of $\alpha$. If $f_{\rm HG}<0.2$, however, this trend is reversed, i.e., $R_0$ increases with increasing $\alpha$. {We note here that two competing factors may affect the resulting $R_0$ due to the change of $\alpha$. On the one hand, a larger $\alpha$ may produce wider post-CE orbits that become more vulnerable to SN kicks and thus lead to the decrease of $R_0$; on the other hand, a larger $\alpha$ can prevent the direct merger of binaries with larger $|E_{\rm bind}|$ (dominated by those with HG donors) at CE stage and thus lead to the increase of $R_0$. When $f_{\rm HG}$ is large, the orbital disruption effect by SN kicks dominates and $R_0$ decreases with increasing $\alpha$, while when $f_{\rm HG}$ is low, the effect by surviving HG donors dominates, and thus the trend of $R_0$ on $\alpha$ reverses.} 
Based on our above results, the latest constraint on the local BNS merger rate given by \citet{2025arXiv250708778A}, i.e., $56_{-40}^{+99}\rm Gpc^{-3}yr^{-1}$
strongly favors models with $f_{\rm HG}>0.8$. 

Third, enforcing all the HG donors to merge with their companion at the CE stage (i.e., $f_{\rm HG}=1$) substantially reduces the {Bayes factor} compared with models with $f_{\rm HG}<1$. This is mainly due to the formation of Galactic BNSs with short orbital periods are significantly suppressed when $f_{\rm HG}=1$. For example, as seen in the middle and lower panels of Figure~\ref{fig:2} for the $P_{\rm orb}-e$ distributions of the mock Galactic BNSs resulting from M16 [$(\alpha,f_{\rm HG})=(5,1)$] and M17 [$(\alpha,f_{\rm HG})=(5,0.8)$], respectively, 
model M17 cannot reproduce those observed Galactic BNSs with $P_{\rm orb}\lesssim 1$ day, while model M16 can nicely match the observations. This can be attributed by the fact that { the absolute binding energy $|E_{\rm bind}|$ of binaries with HG donors entering the CE stage are about $\sim 3$ times larger than those without HG donors, mainly due to their smaller Roche lobe radius. If they are all merged during the CE stage, short periods (e.g., $P_{\rm orb}\lesssim 1$\,day) BNSs cannot be reproduced. However, if part of them can survive after the CE stage, then their relatively larger $|E_{\rm bind}|$  comparing with those in the models with $f_{\rm HG}=1$ at the same stage naturally leads to the formation of BNSs with shorter orbital periods. For example, in model M17, the median orbital period $P_{\rm orb}$  of mock BNS produced from binaries with and without HG donors are about $\sim 0.84$ and $\sim 5.91$ days respectively. }
%$0.84_{-0.40}^{+0.74}$ and $5.91^{+7.43}_{-2.92}$ days respectively. }

At last, we find that only model M11, M16, M17, M19, M21, M22 are compatible with the local merger rate density of BNS mergers measured by GW detection. As for the orbital parameter distribution of galactic BNS systems measured by pulsar radio observation, model M16 gives the highest {Bayes factor} $\log{K_i}=1$ among these 6 models. Therefore, we conclude that  $\alpha=5$ and $f_{\rm HG}=0.8$ is most compatible with current observations. Note that $\alpha=5$ has also been {suggested} by the one dimensional hydrodynamic BNS CE simulation \citep{2019ApJ...883L..45F}.  

\section{Conclusions and Discussions}
\label{sec:con}

In this letter, we
adopt the compact binary population synthesis method to constrain the common envelope (CE) evolution of BNS progenitors, especially the CE ejection efficiency $\alpha$ and the fraction $f_{\rm HG}$ of Hertzsprung gap (HG) donors forced to merge with their companions when entering a CE stage, using the latest estimation of the BNS local merger rate density and the orbital parameter distribution of the observed Galactic BNSs with supplement of a number of newly discovered ones.  
Our main conclusions are summarized as follows.
\begin{itemize}
    \item The CE ejection efficiency $\alpha$ should be larger than $1$, otherwise the observed Galactic BNSs with long orbital period may not be formed. 
    
    \item A fraction of HG donors should not survive from the CE stage, i.e., $f_{\rm HG}>0$, in order to reproduce the comparably low BNS local merger rate density constrained by the preliminary catalog constructed from public alerts of LVK O4 run.
    
    \item  A fraction of HG donors must survive from the CE stage, i.e., $f_{\rm HG}<1$, otherwise the observed Galactic BNSs with short orbital periods cannot be reproduced.

    \item By Bayesian analysis, the BSE model with $\alpha=5$ and $f_{\rm HG}=0.8$ is found to be most compatible with both the constraints of BNS local merger rate density and orbital parameters of observed Galactic BNSs, among all models we simulate.  
    
\end{itemize}

Here we note that there are many complexities one may need to take into account to make a more robust and detailed investigation. First, in this paper, we adopt a $f_{\rm HG}-\rm approximation$ to consider the fraction of HG donors {that} merge with its companion in the CE stage, which is however  oversimplified. In principle, whether a HG donor will merge with their companion star when entering a CE depends on its intrinsic properties such as core mass, envelope mass and radius. Therefore, a detailed hydrodynamic simulation of CE evolution is needed to give a more realistic criteria for the identification of the survival of HG donors in CE. Second, we fix the dispersion of the natal kick as $\sigma_{\rm k,1}=\rm 265~km/s$ for CCSNe and find that $R_0$ may not match with the current constraint if $f_{\rm HG}=0$.  We also examine that when $\sigma_{\rm k,1}=500\rm~km/s $, the value of $R_0$ reduces and may also match with the constraint even when $f_{\rm HG}=1$. However, under this circumstance, the {Bayes factor} of Galactic BNS systems will decreased significantly, since a large fraction of BNS systems with long orbit period are unbounded suffering such high natal kick during the SNe.
Furthermore, \citet{2025arXiv250522102D} re-analyzed the dynamics of pulsars and found that the natal kick velocities tend to be smaller rather than as high as $500\rm~km/s$.
Third, the CE ejection efficiency $\alpha$ actually acts on the orbital shrinkage together with the CE structure parameter $\lambda$ in a manner of $\lambda \alpha$ in Equation~\ref{eq:CE}. Therefore, choosing a different $\lambda$ may also affect the constraints for $\alpha$. In this work, we adopt $\lambda$ from  \citet{2014A&A...563A..83C}, which has been widely used in several CBPS works, such as \textbf{SEVN} \citep{2023MNRAS.524..426I,2025A&A...698A.144S} and \textbf{COMPAS} \citep{2022ApJS..258...34R,2025arXiv250602316M}. But there are also other considerations of $\lambda$, e.g., the fitting formula given by  \citep{2010ApJ...716..114X,2016RAA....16..126W}, which are quite different from that adopted in the this work. We defer a detailed analysis of the effect by the choice of a different $\lambda$ to future work.
In addition, the uncertainty in the cosmic star formation history (SFH) may also affect the $R_0$ estimation \citep[e.g,.][]{2025A&A...698A.144S}. For example, CYL22 find that adopting the SFH extracted from cosmological hydrodynamic simulations, the merger rate density may change by a factor of up to 2, compared with the SFH given by \citet{2014ARA&A..52..415M}, which may affect our constraints.

\section*{Acknowledgements}
We thank the anonymous referee for insightful comments. This work is partly supported by the National Astronomical Observatory of China (grant no. E4TG660101), the Strategic Priority Program of the Chinese Academy of Sciences (grant no. XDB23040100), the Postdoctoral Fellowship Program of CPSF under Grant Number GZB20250735 (ZC), and the National Natural Science Foundation of China under grant nos.\ 12273050.

\section*{Data Availability}

The data underlying this article will be shared on reasonable request to the corresponding author.



\bibliographystyle{mnras}
\bibliography{example} % if your bibtex file is called example.bib

\begin{thebibliography}{}
\makeatletter
\relax
\def\mn@urlcharsother{\let\do\@makeother \do\$\do\&\do\#\do\^\do\_\do\%\do\~}
\def\mn@doi{\begingroup\mn@urlcharsother \@ifnextchar [ {\mn@doi@} {\mn@doi@[]}}
\def\mn@doi@[#1]#2{\def\@tempa{#1}\ifx\@tempa\@empty \href {http://dx.doi.org/#2} {doi:#2}\else \href {http://dx.doi.org/#2} {#1}\fi \endgroup}
\def\mn@eprint#1#2{\mn@eprint@#1:#2::\@nil}
\def\mn@eprint@arXiv#1{\href {http://arxiv.org/abs/#1} {{\tt arXiv:#1}}}
\def\mn@eprint@dblp#1{\href {http://dblp.uni-trier.de/rec/bibtex/#1.xml} {dblp:#1}}
\def\mn@eprint@#1:#2:#3:#4\@nil{\def\@tempa {#1}\def\@tempb {#2}\def\@tempc {#3}\ifx \@tempc \@empty \let \@tempc \@tempb \let \@tempb \@tempa \fi \ifx \@tempb \@empty \def\@tempb {arXiv}\fi \@ifundefined {mn@eprint@\@tempb}{\@tempb:\@tempc}{\expandafter \expandafter \csname mn@eprint@\@tempb\endcsname \expandafter{\@tempc}}}

\bibitem[\protect\citeauthoryear{{Abbott} et~al.,}{{Abbott} et~al.}{2017a}]{2017PhRvL.119p1101A}
{Abbott} B.~P.,  et~al., 2017a, \mn@doi [\prl] {10.1103/PhysRevLett.119.161101}, \href {https://ui.adsabs.harvard.edu/abs/2017PhRvL.119p1101A} {119, 161101}

\bibitem[\protect\citeauthoryear{{Abbott} et~al.,}{{Abbott} et~al.}{2017b}]{2017ApJ...848L..12A}
{Abbott} B.~P.,  et~al., 2017b, \mn@doi [\apjl] {10.3847/2041-8213/aa91c9}, \href {https://ui.adsabs.harvard.edu/abs/2017ApJ...848L..12A} {848, L12}

\bibitem[\protect\citeauthoryear{{Abbott} et~al.,}{{Abbott} et~al.}{2017c}]{2017ApJ...848L..13A}
{Abbott} B.~P.,  et~al., 2017c, \mn@doi [\apjl] {10.3847/2041-8213/aa920c}, \href {https://ui.adsabs.harvard.edu/abs/2017ApJ...848L..13A} {848, L13}

\bibitem[\protect\citeauthoryear{{Abbott} et~al.,}{{Abbott} et~al.}{2019}]{2019PhRvX...9a1001A}
{Abbott} B.~P.,  et~al., 2019, \mn@doi [Physical Review X] {10.1103/PhysRevX.9.011001}, \href {https://ui.adsabs.harvard.edu/abs/2019PhRvX...9a1001A} {9, 011001}

\bibitem[\protect\citeauthoryear{{Aky{\"u}z} et~al.,}{{Aky{\"u}z} et~al.}{2025}]{2025arXiv250708778A}
{Aky{\"u}z} A.,  et~al., 2025, arXiv e-prints, \href {https://ui.adsabs.harvard.edu/abs/2025arXiv250708778A} {p. arXiv:2507.08778}

\bibitem[\protect\citeauthoryear{{Andrews} \& {Mandel}}{{Andrews} \& {Mandel}}{2019}]{2019ApJ...880L...8A}
{Andrews} J.~J.,  {Mandel} I.,  2019, \mn@doi [\apjl] {10.3847/2041-8213/ab2ed1}, \href {https://ui.adsabs.harvard.edu/abs/2019ApJ...880L...8A} {880, L8}

\bibitem[\protect\citeauthoryear{{Andrews} \& {Zezas}}{{Andrews} \& {Zezas}}{2019}]{2019MNRAS.486.3213A}
{Andrews} J.~J.,  {Zezas} A.,  2019, \mn@doi [\mnras] {10.1093/mnras/stz1066}, \href {https://ui.adsabs.harvard.edu/abs/2019MNRAS.486.3213A} {486, 3213}

\bibitem[\protect\citeauthoryear{{Barr} et~al.,}{{Barr} et~al.}{2024}]{Barr+2024Sci...383..275B}
{Barr} E.~D.,  et~al., 2024, \mn@doi [Science] {10.1126/science.adg3005}, \href {https://ui.adsabs.harvard.edu/abs/2024Sci...383..275B} {383, 275}

\bibitem[\protect\citeauthoryear{{Belczynski}, {Taam}, {Kalogera}, {Rasio}  \& {Bulik}}{{Belczynski} et~al.}{2007}]{2007ApJ...662..504B}
{Belczynski} K.,  {Taam} R.~E.,  {Kalogera} V.,  {Rasio} F.~A.,   {Bulik} T.,  2007, \mn@doi [\apj] {10.1086/513562}, \href {https://ui.adsabs.harvard.edu/abs/2007ApJ...662..504B} {662, 504}

\bibitem[\protect\citeauthoryear{{Belczynski}, {Holz}, {Bulik}  \& {O'Shaughnessy}}{{Belczynski} et~al.}{2016a}]{2016Natur.534..512B}
{Belczynski} K.,  {Holz} D.~E.,  {Bulik} T.,   {O'Shaughnessy} R.,  2016a, \mn@doi [\nat] {10.1038/nature18322}, \href {https://ui.adsabs.harvard.edu/abs/2016Natur.534..512B} {534, 512}

\bibitem[\protect\citeauthoryear{{Belczynski}, {Repetto}, {Holz}, {O'Shaughnessy}, {Bulik}, {Berti}, {Fryer}  \& {Dominik}}{{Belczynski} et~al.}{2016b}]{2016ApJ...819..108B}
{Belczynski} K.,  {Repetto} S.,  {Holz} D.~E.,  {O'Shaughnessy} R.,  {Bulik} T.,  {Berti} E.,  {Fryer} C.,   {Dominik} M.,  2016b, \mn@doi [\apj] {10.3847/0004-637X/819/2/108}, \href {https://ui.adsabs.harvard.edu/abs/2016ApJ...819..108B} {819, 108}

\bibitem[\protect\citeauthoryear{{Bhattacharya} \& {van den Heuvel}}{{Bhattacharya} \& {van den Heuvel}}{1991}]{1991PhR...203....1B}
{Bhattacharya} D.,  {van den Heuvel} E.~P.~J.,  1991, \mn@doi [\physrep] {10.1016/0370-1573(91)90064-S}, \href {https://ui.adsabs.harvard.edu/abs/1991PhR...203....1B} {203, 1}

\bibitem[\protect\citeauthoryear{{Bildsten} et~al.,}{{Bildsten} et~al.}{1997}]{1997ApJS..113..367B}
{Bildsten} L.,  et~al., 1997, \mn@doi [\apjs] {10.1086/313060}, \href {https://ui.adsabs.harvard.edu/abs/1997ApJS..113..367B} {113, 367}

\bibitem[\protect\citeauthoryear{{Boesky}, {Broekgaarden}  \& {Berger}}{{Boesky} et~al.}{2024}]{2024ApJ...976...24B}
{Boesky} A.~P.,  {Broekgaarden} F.~S.,   {Berger} E.,  2024, \mn@doi [\apj] {10.3847/1538-4357/ad7fe3}, \href {https://ui.adsabs.harvard.edu/abs/2024ApJ...976...24B} {976, 24}

\bibitem[\protect\citeauthoryear{{Cameron} et~al.,}{{Cameron} et~al.}{2023}]{Cameron+2023MNRAS.523.5064C}
{Cameron} A.~D.,  et~al., 2023, \mn@doi [\mnras] {10.1093/mnras/stad1712}, \href {https://ui.adsabs.harvard.edu/abs/2023MNRAS.523.5064C} {523, 5064}

\bibitem[\protect\citeauthoryear{{Chattopadhyay}, {Stevenson}, {Hurley}, {Rossi}  \& {Flynn}}{{Chattopadhyay} et~al.}{2020}]{2020MNRAS.494.1587C}
{Chattopadhyay} D.,  {Stevenson} S.,  {Hurley} J.~R.,  {Rossi} L.~J.,   {Flynn} C.,  2020, \mn@doi [\mnras] {10.1093/mnras/staa756}, \href {https://ui.adsabs.harvard.edu/abs/2020MNRAS.494.1587C} {494, 1587}

\bibitem[\protect\citeauthoryear{{Chen}, {Lu}, {Wang}, {Jiang}, {Chu}  \& {Ma}}{{Chen} et~al.}{2024}]{2024ApJ...973..159C}
{Chen} Z.,  {Lu} Y.,  {Wang} J.,  {Jiang} Z.,  {Chu} Q.,   {Ma} X.,  2024, \mn@doi [\apj] {10.3847/1538-4357/ad6de0}, \href {https://ui.adsabs.harvard.edu/abs/2024ApJ...973..159C} {973, 159}

\bibitem[\protect\citeauthoryear{{Chu}, {Yu}  \& {Lu}}{{Chu} et~al.}{2022}]{2022MNRAS.509.1557C}
{Chu} Q.,  {Yu} S.,   {Lu} Y.,  2022, \mn@doi [\mnras] {10.1093/mnras/stab2882}, \href {https://ui.adsabs.harvard.edu/abs/2022MNRAS.509.1557C} {509, 1557}

\bibitem[\protect\citeauthoryear{{Chu}, {Lu}  \& {Yu}}{{Chu} et~al.}{2025}]{2025ApJ...980..181C}
{Chu} Q.,  {Lu} Y.,   {Yu} S.,  2025, \mn@doi [\apj] {10.3847/1538-4357/ad90e2}, \href {https://ui.adsabs.harvard.edu/abs/2025ApJ...980..181C} {980, 181}

\bibitem[\protect\citeauthoryear{{Claeys}, {Pols}, {Izzard}, {Vink}  \& {Verbunt}}{{Claeys} et~al.}{2014}]{2014A&A...563A..83C}
{Claeys} J.~S.~W.,  {Pols} O.~R.,  {Izzard} R.~G.,  {Vink} J.,   {Verbunt} F.~W.~M.,  2014, \mn@doi [\aap] {10.1051/0004-6361/201322714}, \href {https://ui.adsabs.harvard.edu/abs/2014A&A...563A..83C} {563, A83}

\bibitem[\protect\citeauthoryear{{Coughlin}, {Dietrich}, {Margalit}  \& {Metzger}}{{Coughlin} et~al.}{2019}]{2019MNRAS.489L..91C}
{Coughlin} M.~W.,  {Dietrich} T.,  {Margalit} B.,   {Metzger} B.~D.,  2019, \mn@doi [\mnras] {10.1093/mnrasl/slz133}, \href {https://ui.adsabs.harvard.edu/abs/2019MNRAS.489L..91C} {489, L91}

\bibitem[\protect\citeauthoryear{{Coulter} et~al.,}{{Coulter} et~al.}{2017}]{2017Sci...358.1556C}
{Coulter} D.~A.,  et~al., 2017, \mn@doi [Science] {10.1126/science.aap9811}, \href {https://ui.adsabs.harvard.edu/abs/2017Sci...358.1556C} {358, 1556}

\bibitem[\protect\citeauthoryear{{Deng}, {Li}, {Shao}  \& {Xu}}{{Deng} et~al.}{2024}]{2024ApJ...963...80D}
{Deng} Z.-L.,  {Li} X.-D.,  {Shao} Y.,   {Xu} K.,  2024, \mn@doi [\apj] {10.3847/1538-4357/ad2357}, \href {https://ui.adsabs.harvard.edu/abs/2024ApJ...963...80D} {963, 80}

\bibitem[\protect\citeauthoryear{{Disberg} \& {Mandel}}{{Disberg} \& {Mandel}}{2025}]{2025arXiv250522102D}
{Disberg} P.,  {Mandel} I.,  2025, \mn@doi [arXiv e-prints] {10.48550/arXiv.2505.22102}, \href {https://ui.adsabs.harvard.edu/abs/2025arXiv250522102D} {p. arXiv:2505.22102}

\bibitem[\protect\citeauthoryear{{Dominik}, {Belczynski}, {Fryer}, {Holz}, {Berti}, {Bulik}, {Mandel}  \& {O'Shaughnessy}}{{Dominik} et~al.}{2012}]{2012ApJ...759...52D}
{Dominik} M.,  {Belczynski} K.,  {Fryer} C.,  {Holz} D.~E.,  {Berti} E.,  {Bulik} T.,  {Mandel} I.,   {O'Shaughnessy} R.,  2012, \mn@doi [\apj] {10.1088/0004-637X/759/1/52}, \href {https://ui.adsabs.harvard.edu/abs/2012ApJ...759...52D} {759, 52}

\bibitem[\protect\citeauthoryear{{Dominik}, {Belczynski}, {Fryer}, {Holz}, {Berti}, {Bulik}, {Mandel}  \& {O'Shaughnessy}}{{Dominik} et~al.}{2013}]{2013ApJ...779...72D}
{Dominik} M.,  {Belczynski} K.,  {Fryer} C.,  {Holz} D.~E.,  {Berti} E.,  {Bulik} T.,  {Mandel} I.,   {O'Shaughnessy} R.,  2013, \mn@doi [\apj] {10.1088/0004-637X/779/1/72}, \href {https://ui.adsabs.harvard.edu/abs/2013ApJ...779...72D} {779, 72}

\bibitem[\protect\citeauthoryear{{Douchin} \& {Haensel}}{{Douchin} \& {Haensel}}{2001}]{2001A&A...380..151D}
{Douchin} F.,  {Haensel} P.,  2001, \mn@doi [\aap] {10.1051/0004-6361:20011402}, \href {https://ui.adsabs.harvard.edu/abs/2001A&A...380..151D} {380, 151}

\bibitem[\protect\citeauthoryear{{Eggleton}, {Fitchett}  \& {Tout}}{{Eggleton} et~al.}{1989}]{1989ApJ...347..998E}
{Eggleton} P.~P.,  {Fitchett} M.~J.,   {Tout} C.~A.,  1989, \mn@doi [\apj] {10.1086/168190}, \href {https://ui.adsabs.harvard.edu/abs/1989ApJ...347..998E} {347, 998}

\bibitem[\protect\citeauthoryear{{Faucher-Gigu{\`e}re} \& {Kaspi}}{{Faucher-Gigu{\`e}re} \& {Kaspi}}{2006}]{2006ApJ...643..332F}
{Faucher-Gigu{\`e}re} C.-A.,  {Kaspi} V.~M.,  2006, \mn@doi [\apj] {10.1086/501516}, \href {https://ui.adsabs.harvard.edu/abs/2006ApJ...643..332F} {643, 332}

\bibitem[\protect\citeauthoryear{{Ferdman} et~al.,}{{Ferdman} et~al.}{2014}]{Ferdman+2014MNRAS.443.2183F}
{Ferdman} R.~D.,  et~al., 2014, \mn@doi [\mnras] {10.1093/mnras/stu1223}, \href {https://ui.adsabs.harvard.edu/abs/2014MNRAS.443.2183F} {443, 2183}

\bibitem[\protect\citeauthoryear{{Ferreras}, {Wyse}  \& {Silk}}{{Ferreras} et~al.}{2003}]{2003MNRAS.345.1381F}
{Ferreras} I.,  {Wyse} R. F.~G.,   {Silk} J.,  2003, \mn@doi [\mnras] {10.1046/j.1365-2966.2003.07056.x}, \href {https://ui.adsabs.harvard.edu/abs/2003MNRAS.345.1381F} {345, 1381}

\bibitem[\protect\citeauthoryear{{Fonseca}, {Stairs}  \& {Thorsett}}{{Fonseca} et~al.}{2014}]{Fonseca+2014ApJ...787...82F}
{Fonseca} E.,  {Stairs} I.~H.,   {Thorsett} S.~E.,  2014, \mn@doi [\apj] {10.1088/0004-637X/787/1/82}, \href {https://ui.adsabs.harvard.edu/abs/2014ApJ...787...82F} {787, 82}

\bibitem[\protect\citeauthoryear{{Fragione}, {Grishin}, {Leigh}, {Perets}  \& {Perna}}{{Fragione} et~al.}{2019}]{2019MNRAS.488...47F}
{Fragione} G.,  {Grishin} E.,  {Leigh} N. W.~C.,  {Perets} H.~B.,   {Perna} R.,  2019, \mn@doi [\mnras] {10.1093/mnras/stz1651}, \href {https://ui.adsabs.harvard.edu/abs/2019MNRAS.488...47F} {488, 47}

\bibitem[\protect\citeauthoryear{{Fragos}, {Andrews}, {Ramirez-Ruiz}, {Meynet}, {Kalogera}, {Taam}  \& {Zezas}}{{Fragos} et~al.}{2019}]{2019ApJ...883L..45F}
{Fragos} T.,  {Andrews} J.~J.,  {Ramirez-Ruiz} E.,  {Meynet} G.,  {Kalogera} V.,  {Taam} R.~E.,   {Zezas} A.,  2019, \mn@doi [\apjl] {10.3847/2041-8213/ab40d1}, \href {https://ui.adsabs.harvard.edu/abs/2019ApJ...883L..45F} {883, L45}

\bibitem[\protect\citeauthoryear{{Fragos} et~al.,}{{Fragos} et~al.}{2023}]{2023ApJS..264...45F}
{Fragos} T.,  et~al., 2023, \mn@doi [\apjs] {10.3847/1538-4365/ac90c1}, \href {https://ui.adsabs.harvard.edu/abs/2023ApJS..264...45F} {264, 45}

\bibitem[\protect\citeauthoryear{{Fryer}, {Belczynski}, {Wiktorowicz}, {Dominik}, {Kalogera}  \& {Holz}}{{Fryer} et~al.}{2012}]{2012ApJ...749...91F}
{Fryer} C.~L.,  {Belczynski} K.,  {Wiktorowicz} G.,  {Dominik} M.,  {Kalogera} V.,   {Holz} D.~E.,  2012, \mn@doi [\apj] {10.1088/0004-637X/749/1/91}, \href {https://ui.adsabs.harvard.edu/abs/2012ApJ...749...91F} {749, 91}

\bibitem[\protect\citeauthoryear{{Giacobbo} \& {Mapelli}}{{Giacobbo} \& {Mapelli}}{2018}]{2018MNRAS.480.2011G}
{Giacobbo} N.,  {Mapelli} M.,  2018, \mn@doi [\mnras] {10.1093/mnras/sty1999}, \href {https://ui.adsabs.harvard.edu/abs/2018MNRAS.480.2011G} {480, 2011}

\bibitem[\protect\citeauthoryear{{Han}}{{Han}}{1998}]{1998MNRAS.296.1019H}
{Han} Z.,  1998, \mn@doi [\mnras] {10.1046/j.1365-8711.1998.01475.x}, \href {https://ui.adsabs.harvard.edu/abs/1998MNRAS.296.1019H} {296, 1019}

\bibitem[\protect\citeauthoryear{{Haniewicz}, {Ferdman}, {Freire}, {Champion}, {Bunting}, {Lorimer}  \& {McLaughlin}}{{Haniewicz} et~al.}{2021}]{Haniewicz+2021MNRAS.500.4620H}
{Haniewicz} H.~T.,  {Ferdman} R.~D.,  {Freire} P.~C.~C.,  {Champion} D.~J.,  {Bunting} K.~A.,  {Lorimer} D.~R.,   {McLaughlin} M.~A.,  2021, \mn@doi [\mnras] {10.1093/mnras/staa3466}, \href {https://ui.adsabs.harvard.edu/abs/2021MNRAS.500.4620H} {500, 4620}

\bibitem[\protect\citeauthoryear{{Hobbs}, {Lorimer}, {Lyne}  \& {Kramer}}{{Hobbs} et~al.}{2005}]{2005MNRAS.360..974H}
{Hobbs} G.,  {Lorimer} D.~R.,  {Lyne} A.~G.,   {Kramer} M.,  2005, \mn@doi [\mnras] {10.1111/j.1365-2966.2005.09087.x}, \href {https://ui.adsabs.harvard.edu/abs/2005MNRAS.360..974H} {360, 974}

\bibitem[\protect\citeauthoryear{{Hulse} \& {Taylor}}{{Hulse} \& {Taylor}}{1975}]{1975ApJ...195L..51H}
{Hulse} R.~A.,  {Taylor} J.~H.,  1975, \mn@doi [\apjl] {10.1086/181708}, \href {https://ui.adsabs.harvard.edu/abs/1975ApJ...195L..51H} {195, L51}

\bibitem[\protect\citeauthoryear{{Hurley}, {Pols}  \& {Tout}}{{Hurley} et~al.}{2000}]{2000MNRAS.315..543H}
{Hurley} J.~R.,  {Pols} O.~R.,   {Tout} C.~A.,  2000, \mn@doi [\mnras] {10.1046/j.1365-8711.2000.03426.x}, \href {https://ui.adsabs.harvard.edu/abs/2000MNRAS.315..543H} {315, 543}

\bibitem[\protect\citeauthoryear{{Hurley}, {Tout}  \& {Pols}}{{Hurley} et~al.}{2002}]{2002MNRAS.329..897H}
{Hurley} J.~R.,  {Tout} C.~A.,   {Pols} O.~R.,  2002, \mn@doi [\mnras] {10.1046/j.1365-8711.2002.05038.x}, \href {https://ui.adsabs.harvard.edu/abs/2002MNRAS.329..897H} {329, 897}

\bibitem[\protect\citeauthoryear{{Iorio} et~al.,}{{Iorio} et~al.}{2023}]{2023MNRAS.524..426I}
{Iorio} G.,  et~al., 2023, \mn@doi [\mnras] {10.1093/mnras/stad1630}, \href {https://ui.adsabs.harvard.edu/abs/2023MNRAS.524..426I} {524, 426}

\bibitem[\protect\citeauthoryear{{Ivanova} \& {Taam}}{{Ivanova} \& {Taam}}{2004}]{2004ApJ...601.1058I}
{Ivanova} N.,  {Taam} R.~E.,  2004, \mn@doi [\apj] {10.1086/380561}, \href {https://ui.adsabs.harvard.edu/abs/2004ApJ...601.1058I} {601, 1058}

\bibitem[\protect\citeauthoryear{{Jacoby}, {Cameron}, {Jenet}, {Anderson}, {Murty}  \& {Kulkarni}}{{Jacoby} et~al.}{2006}]{Jacoby+2006ApJ...644L.113J}
{Jacoby} B.~A.,  {Cameron} P.~B.,  {Jenet} F.~A.,  {Anderson} S.~B.,  {Murty} R.~N.,   {Kulkarni} S.~R.,  2006, \mn@doi [\apjl] {10.1086/505742}, \href {https://ui.adsabs.harvard.edu/abs/2006ApJ...644L.113J} {644, L113}

\bibitem[\protect\citeauthoryear{{Janssen}, {Stappers}, {Kramer}, {Nice}, {Jessner}, {Cognard}  \& {Purver}}{{Janssen} et~al.}{2008}]{Janssen+2008AA...490..753J}
{Janssen} G.~H.,  {Stappers} B.~W.,  {Kramer} M.,  {Nice} D.~J.,  {Jessner} A.,  {Cognard} I.,   {Purver} M.~B.,  2008, \mn@doi [\aap] {10.1051/0004-6361:200810076}, \href {https://ui.adsabs.harvard.edu/abs/2008A&A...490..753J} {490, 753}

\bibitem[\protect\citeauthoryear{{Keith}, {Kramer}, {Lyne}, {Eatough}, {Stairs}, {Possenti}, {Camilo}  \& {Manchester}}{{Keith} et~al.}{2009}]{Keith+2009MNRAS.393..623K}
{Keith} M.~J.,  {Kramer} M.,  {Lyne} A.~G.,  {Eatough} R.~P.,  {Stairs} I.~H.,  {Possenti} A.,  {Camilo} F.,   {Manchester} R.~N.,  2009, \mn@doi [\mnras] {10.1111/j.1365-2966.2008.14234.x}, \href {https://ui.adsabs.harvard.edu/abs/2009MNRAS.393..623K} {393, 623}

\bibitem[\protect\citeauthoryear{{Kramer} et~al.,}{{Kramer} et~al.}{2021}]{Kramer+2021PhRvX..11d1050K}
{Kramer} M.,  et~al., 2021, \mn@doi [PhRvX] {10.1103/PhysRevX.11.041050}, \href {https://ui.adsabs.harvard.edu/abs/2021PhRvX..11d1050K} {11, 041050}

\bibitem[\protect\citeauthoryear{{Kroupa}}{{Kroupa}}{2001}]{2001MNRAS.322..231K}
{Kroupa} P.,  2001, \mn@doi [\mnras] {10.1046/j.1365-8711.2001.04022.x}, \href {https://ui.adsabs.harvard.edu/abs/2001MNRAS.322..231K} {322, 231}

\bibitem[\protect\citeauthoryear{{Lazarus} et~al.,}{{Lazarus} et~al.}{2016}]{Lazarus+2016ApJ...831..150L}
{Lazarus} P.,  et~al., 2016, \mn@doi [\apj] {10.3847/0004-637X/831/2/150}, \href {https://ui.adsabs.harvard.edu/abs/2016ApJ...831..150L} {831, 150}

\bibitem[\protect\citeauthoryear{{Madau} \& {Dickinson}}{{Madau} \& {Dickinson}}{2014}]{2014ARA&A..52..415M}
{Madau} P.,  {Dickinson} M.,  2014, \mn@doi [\araa] {10.1146/annurev-astro-081811-125615}, \href {https://ui.adsabs.harvard.edu/abs/2014ARA&A..52..415M} {52, 415}

\bibitem[\protect\citeauthoryear{{Mandel} \& {Broekgaarden}}{{Mandel} \& {Broekgaarden}}{2022}]{2022LRR....25....1M}
{Mandel} I.,  {Broekgaarden} F.~S.,  2022, \mn@doi [Living Reviews in Relativity] {10.1007/s41114-021-00034-3}, \href {https://ui.adsabs.harvard.edu/abs/2022LRR....25....1M} {25, 1}

\bibitem[\protect\citeauthoryear{{Mandel} et~al.,}{{Mandel} et~al.}{2025}]{2025arXiv250602316M}
{Mandel} I.,  et~al., 2025, \mn@doi [arXiv e-prints] {10.48550/arXiv.2506.02316}, \href {https://ui.adsabs.harvard.edu/abs/2025arXiv250602316M} {p. arXiv:2506.02316}

\bibitem[\protect\citeauthoryear{{Mapelli}, {Giacobbo}, {Ripamonti}  \& {Spera}}{{Mapelli} et~al.}{2017}]{2017MNRAS.472.2422M}
{Mapelli} M.,  {Giacobbo} N.,  {Ripamonti} E.,   {Spera} M.,  2017, \mn@doi [\mnras] {10.1093/mnras/stx2123}, \href {https://ui.adsabs.harvard.edu/abs/2017MNRAS.472.2422M} {472, 2422}

\bibitem[\protect\citeauthoryear{{Marchant}, {Pappas}, {Gallegos-Garcia}, {Berry}, {Taam}, {Kalogera}  \& {Podsiadlowski}}{{Marchant} et~al.}{2021}]{2021A&A...650A.107M}
{Marchant} P.,  {Pappas} K. M.~W.,  {Gallegos-Garcia} M.,  {Berry} C. P.~L.,  {Taam} R.~E.,  {Kalogera} V.,   {Podsiadlowski} P.,  2021, \mn@doi [\aap] {10.1051/0004-6361/202039992}, \href {https://ui.adsabs.harvard.edu/abs/2021A&A...650A.107M} {650, A107}

\bibitem[\protect\citeauthoryear{{Martinez} et~al.,}{{Martinez} et~al.}{2015}]{Martinez+2015ApJ...812..143M}
{Martinez} J.~G.,  et~al., 2015, \mn@doi [\apj] {10.1088/0004-637X/812/2/143}, \href {https://ui.adsabs.harvard.edu/abs/2015ApJ...812..143M} {812, 143}

\bibitem[\protect\citeauthoryear{{Martinez} et~al.,}{{Martinez} et~al.}{2017}]{Martinez+2017ApJ...851L..29M}
{Martinez} J.~G.,  et~al., 2017, \mn@doi [\apjl] {10.3847/2041-8213/aa9d87}, \href {https://ui.adsabs.harvard.edu/abs/2017ApJ...851L..29M} {851, L29}

\bibitem[\protect\citeauthoryear{{McEwen} et~al.,}{{McEwen} et~al.}{2024}]{McEwen+2024ApJ...962..167M}
{McEwen} A.~E.,  et~al., 2024, \mn@doi [\apj] {10.3847/1538-4357/ad11f0}, \href {https://ui.adsabs.harvard.edu/abs/2024ApJ...962..167M} {962, 167}

\bibitem[\protect\citeauthoryear{{Miyaji}, {Nomoto}, {Yokoi}  \& {Sugimoto}}{{Miyaji} et~al.}{1980}]{1980PASJ...32..303M}
{Miyaji} S.,  {Nomoto} K.,  {Yokoi} K.,   {Sugimoto} D.,  1980, \pasj, \href {https://ui.adsabs.harvard.edu/abs/1980PASJ...32..303M} {32, 303}

\bibitem[\protect\citeauthoryear{{Ng} et~al.,}{{Ng} et~al.}{2018}]{2018MNRAS.476.4315N}
{Ng} C.,  et~al., 2018, \mn@doi [\mnras] {10.1093/mnras/sty482}, \href {https://ui.adsabs.harvard.edu/abs/2018MNRAS.476.4315N} {476, 4315}

\bibitem[\protect\citeauthoryear{{Olausen} \& {Kaspi}}{{Olausen} \& {Kaspi}}{2014}]{2014ApJS..212....6O}
{Olausen} S.~A.,  {Kaspi} V.~M.,  2014, \mn@doi [\apjs] {10.1088/0067-0049/212/1/6}, \href {https://ui.adsabs.harvard.edu/abs/2014ApJS..212....6O} {212, 6}

\bibitem[\protect\citeauthoryear{{Ridley} \& {Lorimer}}{{Ridley} \& {Lorimer}}{2010}]{2010MNRAS.404.1081R}
{Ridley} J.~P.,  {Lorimer} D.~R.,  2010, \mn@doi [\mnras] {10.1111/j.1365-2966.2010.16342.x}, \href {https://ui.adsabs.harvard.edu/abs/2010MNRAS.404.1081R} {404, 1081}

\bibitem[\protect\citeauthoryear{{Ridolfi}, {Freire}, {Gupta}  \& {Ransom}}{{Ridolfi} et~al.}{2019}]{Ridolfi+2019MNRAS.490.3860R}
{Ridolfi} A.,  {Freire} P.~C.~C.,  {Gupta} Y.,   {Ransom} S.~M.,  2019, \mn@doi [\mnras] {10.1093/mnras/stz2645}, \href {https://ui.adsabs.harvard.edu/abs/2019MNRAS.490.3860R} {490, 3860}

\bibitem[\protect\citeauthoryear{{Riley} et~al.,}{{Riley} et~al.}{2022}]{2022ApJS..258...34R}
{Riley} J.,  et~al., 2022, \mn@doi [\apjs] {10.3847/1538-4365/ac416c}, \href {https://ui.adsabs.harvard.edu/abs/2022ApJS..258...34R} {258, 34}

\bibitem[\protect\citeauthoryear{{R{\"o}pke} \& {De Marco}}{{R{\"o}pke} \& {De Marco}}{2023}]{2023LRCA....9....2R}
{R{\"o}pke} F.~K.,  {De Marco} O.,  2023, \mn@doi [Living Reviews in Computational Astrophysics] {10.1007/s41115-023-00017-x}, \href {https://ui.adsabs.harvard.edu/abs/2023LRCA....9....2R} {9, 2}

\bibitem[\protect\citeauthoryear{{Santoliquido}, {Mapelli}, {Bouffanais}, {Giacobbo}, {Di Carlo}, {Rastello}, {Artale}  \& {Ballone}}{{Santoliquido} et~al.}{2020}]{2020ApJ...898..152S}
{Santoliquido} F.,  {Mapelli} M.,  {Bouffanais} Y.,  {Giacobbo} N.,  {Di Carlo} U.~N.,  {Rastello} S.,  {Artale} M.~C.,   {Ballone} A.,  2020, \mn@doi [\apj] {10.3847/1538-4357/ab9b78}, \href {https://ui.adsabs.harvard.edu/abs/2020ApJ...898..152S} {898, 152}

\bibitem[\protect\citeauthoryear{{Sengar} et~al.,}{{Sengar} et~al.}{2022}]{Sengar+2022MNRAS.512.5782S}
{Sengar} R.,  et~al., 2022, \mn@doi [\mnras] {10.1093/mnras/stac821}, \href {https://ui.adsabs.harvard.edu/abs/2022MNRAS.512.5782S} {512, 5782}

\bibitem[\protect\citeauthoryear{{Sgalletta} et~al.,}{{Sgalletta} et~al.}{2023}]{2023MNRAS.526.2210S}
{Sgalletta} C.,  et~al., 2023, \mn@doi [\mnras] {10.1093/mnras/stad2768}, \href {https://ui.adsabs.harvard.edu/abs/2023MNRAS.526.2210S} {526, 2210}

\bibitem[\protect\citeauthoryear{{Sgalletta}, {Mapelli}, {Boco}, {Santoliquido}, {Artale}, {Iorio}, {Lapi}  \& {Spera}}{{Sgalletta} et~al.}{2025}]{2025A&A...698A.144S}
{Sgalletta} C.,  {Mapelli} M.,  {Boco} L.,  {Santoliquido} F.,  {Artale} M.~C.,  {Iorio} G.,  {Lapi} A.,   {Spera} M.,  2025, \mn@doi [\aap] {10.1051/0004-6361/202452757}, \href {https://ui.adsabs.harvard.edu/abs/2025A&A...698A.144S} {698, A144}

\bibitem[\protect\citeauthoryear{{Snaith}, {Haywood}, {Di Matteo}, {Lehnert}, {Combes}, {Katz}  \& {G{\'o}mez}}{{Snaith} et~al.}{2014}]{2014ApJ...781L..31S}
{Snaith} O.~N.,  {Haywood} M.,  {Di Matteo} P.,  {Lehnert} M.~D.,  {Combes} F.,  {Katz} D.,   {G{\'o}mez} A.,  2014, \mn@doi [\apjl] {10.1088/2041-8205/781/2/L31}, \href {https://ui.adsabs.harvard.edu/abs/2014ApJ...781L..31S} {781, L31}

\bibitem[\protect\citeauthoryear{{Stovall} et~al.,}{{Stovall} et~al.}{2018}]{Stovall+2018ApJ...854L..22S}
{Stovall} K.,  et~al., 2018, \mn@doi [ApJL] {10.3847/2041-8213/aaad06}, \href {https://ui.adsabs.harvard.edu/abs/2018ApJ...854L..22S} {854, L22}

\bibitem[\protect\citeauthoryear{{Su} et~al.,}{{Su} et~al.}{2024a}]{2024MNRAS.530.1506S}
{Su} W.~Q.,  et~al., 2024a, \mn@doi [\mnras] {10.1093/mnras/stae888}, \href {https://ui.adsabs.harvard.edu/abs/2024MNRAS.530.1506S} {530, 1506}

\bibitem[\protect\citeauthoryear{{Su} et~al.,}{{Su} et~al.}{2024b}]{Su+2024MNRAS.530.1506S}
{Su} W.~Q.,  et~al., 2024b, \mn@doi [\mnras] {10.1093/mnras/stae888}, \href {https://ui.adsabs.harvard.edu/abs/2024MNRAS.530.1506S} {530, 1506}

\bibitem[\protect\citeauthoryear{{Swiggum} et~al.,}{{Swiggum} et~al.}{2015}]{Swiggum+2015ApJ...805..156S}
{Swiggum} J.~K.,  et~al., 2015, \mn@doi [\apj] {10.1088/0004-637X/805/2/156}, \href {https://ui.adsabs.harvard.edu/abs/2015ApJ...805..156S} {805, 156}

\bibitem[\protect\citeauthoryear{{Swiggum} et~al.,}{{Swiggum} et~al.}{2023}]{Swiggum+2023ApJ...944..154S}
{Swiggum} J.~K.,  et~al., 2023, \mn@doi [\apj] {10.3847/1538-4357/acb43f}, \href {https://ui.adsabs.harvard.edu/abs/2023ApJ...944..154S} {944, 154}

\bibitem[\protect\citeauthoryear{{Tauris}, {Langer}  \& {Podsiadlowski}}{{Tauris} et~al.}{2015}]{2015MNRAS.451.2123T}
{Tauris} T.~M.,  {Langer} N.,   {Podsiadlowski} P.,  2015, \mn@doi [\mnras] {10.1093/mnras/stv990}, \href {https://ui.adsabs.harvard.edu/abs/2015MNRAS.451.2123T} {451, 2123}

\bibitem[\protect\citeauthoryear{{The LIGO Scientific Collaboration} et~al.,}{{The LIGO Scientific Collaboration} et~al.}{2021}]{2021arXiv211103606T}
{The LIGO Scientific Collaboration} et~al., 2021, arXiv e-prints, \href {https://ui.adsabs.harvard.edu/abs/2021arXiv211103606T} {p. arXiv:2111.03606}

\bibitem[\protect\citeauthoryear{{Vigna-G{\'o}mez} et~al.,}{{Vigna-G{\'o}mez} et~al.}{2018}]{2018MNRAS.481.4009V}
{Vigna-G{\'o}mez} A.,  et~al., 2018, \mn@doi [\mnras] {10.1093/mnras/sty2463}, \href {https://ui.adsabs.harvard.edu/abs/2018MNRAS.481.4009V} {481, 4009}

\bibitem[\protect\citeauthoryear{{Wang}, {Jia}  \& {Li}}{{Wang} et~al.}{2016}]{2016RAA....16..126W}
{Wang} C.,  {Jia} K.,   {Li} X.-D.,  2016, \mn@doi [Research in Astronomy and Astrophysics] {10.1088/1674-4527/16/8/126}, \href {https://ui.adsabs.harvard.edu/abs/2016RAA....16..126W} {16, 126}

\bibitem[\protect\citeauthoryear{{Wang} et~al.,}{{Wang} et~al.}{2025a}]{2025RAA....25a4003W}
{Wang} P.~F.,  et~al., 2025a, \mn@doi [Research in Astronomy and Astrophysics] {10.1088/1674-4527/ada3b8}, \href {https://ui.adsabs.harvard.edu/abs/2025RAA....25a4003W} {25, 014003}

\bibitem[\protect\citeauthoryear{{Wang} et~al.,}{{Wang} et~al.}{2025b}]{Wang+2024arXiv241203062W}
{Wang} P.~F.,  et~al., 2025b, \mn@doi [Research in Astronomy and Astrophysics] {10.1088/1674-4527/ada3b8}, \href {https://ui.adsabs.harvard.edu/abs/2025RAA....25a4003W} {25, 014003}

\bibitem[\protect\citeauthoryear{{Webbink}}{{Webbink}}{1984}]{1984ApJ...277..355W}
{Webbink} R.~F.,  1984, \mn@doi [\apj] {10.1086/161701}, \href {https://ui.adsabs.harvard.edu/abs/1984ApJ...277..355W} {277, 355}

\bibitem[\protect\citeauthoryear{{Weisberg} \& {Huang}}{{Weisberg} \& {Huang}}{2016}]{Weisberg+2016ApJ...829...55W}
{Weisberg} J.~M.,  {Huang} Y.,  2016, \mn@doi [\apj] {10.3847/0004-637X/829/1/55}, \href {https://ui.adsabs.harvard.edu/abs/2016ApJ...829...55W} {829, 55}

\bibitem[\protect\citeauthoryear{{Wu} et~al.,}{{Wu} et~al.}{2023}]{Wu+2023ApJ...958L..17W}
{Wu} Q.~D.,  et~al., 2023, \mn@doi [\apjl] {10.3847/2041-8213/ad0887}, \href {https://ui.adsabs.harvard.edu/abs/2023ApJ...958L..17W} {958, L17}

\bibitem[\protect\citeauthoryear{{Xu} \& {Li}}{{Xu} \& {Li}}{2010}]{2010ApJ...716..114X}
{Xu} X.-J.,  {Li} X.-D.,  2010, \mn@doi [\apj] {10.1088/0004-637X/716/1/114}, \href {https://ui.adsabs.harvard.edu/abs/2010ApJ...716..114X} {716, 114}

\bibitem[\protect\citeauthoryear{{Ye}, {Fong}, {Kremer}, {Rodriguez}, {Chatterjee}, {Fragione}  \& {Rasio}}{{Ye} et~al.}{2020}]{2020ApJ...888L..10Y}
{Ye} C.~S.,  {Fong} W.-f.,  {Kremer} K.,  {Rodriguez} C.~L.,  {Chatterjee} S.,  {Fragione} G.,   {Rasio} F.~A.,  2020, \mn@doi [\apjl] {10.3847/2041-8213/ab5dc5}, \href {https://ui.adsabs.harvard.edu/abs/2020ApJ...888L..10Y} {888, L10}

\bibitem[\protect\citeauthoryear{{Yu} \& {Jeffery}}{{Yu} \& {Jeffery}}{2010}]{2010A&A...521A..85Y}
{Yu} S.,  {Jeffery} C.~S.,  2010, \mn@doi [\aap] {10.1051/0004-6361/201014827}, \href {https://ui.adsabs.harvard.edu/abs/2010A&A...521A..85Y} {521, A85}

\bibitem[\protect\citeauthoryear{{van Leeuwen} et~al.,}{{van Leeuwen} et~al.}{2015}]{Leeuwen+2015ApJ...798..118V}
{van Leeuwen} J.,  et~al., 2015, \mn@doi [\apj] {10.1088/0004-637X/798/2/118}, \href {https://ui.adsabs.harvard.edu/abs/2015ApJ...798..118V} {798, 118}

\makeatother
\end{thebibliography}

%\begin{thebibliography}{99}
%\end{thebibliography}

%%%%%%%%%%%%%%%%%%%%%%%%%%%%%%%%%%%%%%%%%%%%%%%%%%

%%%%%%%%%%%%%%%%% APPENDICES %%%%%%%%%%%%%%%%%%%%%

\appendix

\begin{table}
    %\centering
    \label{tab:BNS}
    \caption{Orbital parameters of known Galactic BNS systems in Milky way field observed by pulsar radio telescopes. The 2nd column shows their names ( candidates with $\ast$). The 3rd and 4th columns show their orbital period $P_{\rm orb}$ and eccentricity $e$. The references for each system are listed in the 5th column. {Notably, pulsar J0737-3039A and B are in the same BNS as a double pulsar system denoted as index 23. }}
    %\lyj{are 23 and 24 actually the same system, with exactly the same parameters???} \czw{Yes, it is a double pulsar system. }}
    \fontsize{8}{12}\selectfont % 紧凑字体
    \begin{tabular}{lrrrl}
    \hline
    No. & PSR & $P_{\rm orb}$ (day) & $e$ & Ref. \\
    \hline
1  & J0514$-$4002A   & 18.7852  & 0.8880 & \cite{Ridolfi+2019MNRAS.490.3860R} \\
2  & J0514$-$4002E   & 7.4479   & 0.7079 & \cite{Barr+2024Sci...383..275B} \\
3  & B2127+11C       & 0.3353   & 0.6814 & \cite{Jacoby+2006ApJ...644L.113J} \\
4  & B1913+16        & 0.3230   & 0.6171 & \cite{Weisberg+2016ApJ...829...55W} \\
5  & J1757$-$1854    & 0.1835   & 0.6058 & \cite{Cameron+2023MNRAS.523.5064C} \\
6  & J2150+3427      & 10.5921  & 0.6015 & \cite{Wu+2023ApJ...958L..17W} \\
7  & J0509+3801      & 0.3796   & 0.5864 & \cite{McEwen+2024ApJ...962..167M} \\
8 & J1930$-$1852    & 45.0600  & 0.3989 & \cite{Swiggum+2015ApJ...805..156S} \\
9 & J1901+0658      & 14.4548  & 0.3662 & \cite{Su+2024MNRAS.530.1506S} \\
10 & J1759+5036$^{\ast}$      & 2.0430   & 0.3083 & \cite{McEwen+2024ApJ...962..167M} \\
11 & J1753$-$2240$^{\ast}$    & 13.6376  & 0.3036 & \cite{Keith+2009MNRAS.393..623K} \\
12 & J0528+3529      & 11.7262  & 0.2901 & \cite{Wang+2024arXiv241203062W} \\
13 & B1534+12        & 0.4207   & 0.2737 & \cite{Fonseca+2014ApJ...787...82F} \\
14 & J1518+4904      & 8.6340   & 0.2495 & \cite{Janssen+2008AA...490..753J} \\
15 & J1844$-$0128    & 10.6003  & 0.2349 & \cite{Wang+2024arXiv241203062W} \\
16 & J1018$-$1523$^{\ast}$    & 8.9840   & 0.2277 & \cite{Swiggum+2023ApJ...944..154S} \\
17 & J1756$-$2251    & 0.3196   & 0.1806 & \cite{Ferdman+2014MNRAS.443.2183F} \\
18 & J1411+2551      & 2.6159   & 0.1699 & \cite{Martinez+2017ApJ...851L..29M} \\
19 & J1829+2456      & 1.1760   & 0.1391 & \cite{Haniewicz+2021MNRAS.500.4620H} \\
20 & J0453+1559      & 4.0725   & 0.1125 & \cite{Martinez+2015ApJ...812..143M} \\
21 & J1913+1102      & 0.2063   & 0.08954 & \cite{Lazarus+2016ApJ...831..150L} \\
22 & J1755$-$2550$^{\ast}$    & 9.6963   & 0.08935 & \cite{2018MNRAS.476.4315N} \\
23 & J0737$-$3039B   & 0.1023   & 0.08778 & \cite{Kramer+2021PhRvX..11d1050K} \\
23 & J0737$-$3039A   & 0.1023   & 0.08778 & \cite{Kramer+2021PhRvX..11d1050K} \\
24 & J1906+0746$^{\ast}$      & 0.1660   & 0.08530 & \cite{Leeuwen+2015ApJ...798..118V} \\
25 & J1325$-$6253    & 1.8156   & 0.06401 & \cite{Sengar+2022MNRAS.512.5782S} \\
26 & J1946+2052      & 0.0785   & 0.06385 & \cite{Stovall+2018ApJ...854L..22S} \\
    \hline
    \end{tabular}
    \label{tab:DNS_params_ordered}
\end{table}

\begin{table}
\fontsize{8}{12}\selectfont %

\caption{The 2nd and 3rd columns show the parameter settings for different BSE models, corresponding to model from M1 to M22 in the 1st column. The 4th column shows the resulted local merger rate density of BNS systems $R_0$. The 5th and 6th column shows the relative {Bayes factor} $\log K$,obtained by matching the detected Galactic BNS systems with and without 5 candidates in MW field shown in Table~\ref{tab:BNS} respectively, chosen model 6 as the reference model. }
\label{table:para} 
\centering
\begin{tabular}{lccccccc} 
\hline
Model & $\alpha$ &$f_{\rm HG}$ & $R_{0}[\rm Gpc^{-3} \rm yr^{-1}]$ & $\log K_i$& $\log K_i$(Remove candidate)
\\
\hline

M1  & 1.0 & 0.0 & 373 & -12 & -8 \\
M2  & 1.0 & 0.2 & 361 & -12 & -8 \\
M3  & 1.0 & 0.4 & 354 & -12 & -8 \\
M4  & 1.0 & 0.6 & 351 & -10 & -7 \\
M5  & 1.0 & 0.8 & 331 & -10 & -8 \\
M6  & 3.0 & 0.0 & 383 &   0 &  0 \\
M7  & 3.0 & 0.2 & 345 &   1 &  1 \\
M8  & 3.0 & 0.4 & 293 &   1 &  1 \\
M9  & 3.0 & 0.6 & 249 &   3 &  1 \\
M10 & 3.0 & 0.8 & 209 &   2 &  0 \\
M11 & 3.0 & 1.0 & 155 & -11 & -12\\
M12 & 5.0 & 0.0 & 445 &   5 &  2 \\
M13 & 5.0 & 0.2 & 368 &   3 &  1 \\
M14 & 5.0 & 0.4 & 285 &   2 &  1 \\
M15 & 5.0 & 0.6 & 236 &   2 &  0 \\
M16 & 5.0 & 0.8 & 149 &   1 & -1 \\
M17 & 5.0 & 1.0 &  79 & -18 & -20\\
M18 & 7.0 & 0.6 & 172 &   1 &  0 \\
M19 & 7.0 & 0.8 & 112 &  -1 &  -2\\
M20 & 10 & 0.4 & 174 &  2 &  0 \\
M21& 10 & 0.6 & 123 & -1 &  -2 \\
M22 & 10 & 0.8 &  76 & -5 & -7 \\
\hline
\end{tabular}
\end{table}

%%%%%%%%%%%%%%%%%%%%%%%%%%%%%%%%%%%%%%%%%%%%%%%%%%

% Don't change these lines
\bsp	% typesetting comment
\label{lastpage}
\end{document}